\documentclass[conference]{IEEEtran}
\IEEEoverridecommandlockouts
\usepackage{cite}
\usepackage{orcidlink}
\usepackage{amsmath,amssymb,amsfonts}
\usepackage{algorithmic}
\usepackage{graphicx}
\usepackage{textcomp}
\usepackage{xcolor}
\usepackage{braket}
\usepackage{multirow}
\usepackage{hyperref}
\usepackage{fontawesome}
\usepackage[authormarkuptext=name,commentmarkup=uwave]{changes}
\definechangesauthor[name={JRH},color=green!70!black]{jonte}

\definechangesauthor[name={NC},color=magenta]{nick}

\definechangesauthor[name={SVP},color=cyan]{soraya}

\def\BibTeX{{\rm B\kern-.05em{\sc i\kern-.025em b}\kern-.08em
    T\kern-.1667em\lower.7ex\hbox{E}\kern-.125emX}}
\begin{document}
\title{Impact of Data Loss in Postprocessing on Training and Inference of Quantum Neural Networks}
\author{
Soraya V Panambalom$^{1,*}$\orcidlink{0009-0009-2827-7988}, Edoardo Altamura$^{2,3}$\orcidlink{0000-0001-6973-1897}, Nick Chancellor$^1$\orcidlink{0000-0002-1293-0761}, and Jonte R Hance$^{1,\dagger}$\orcidlink{0000-0001-8587-7618}
\thanks{\textit{$^1$Quantum Group, School of Computing, Newcastle University, 1 Science Square, Newcastle upon Tyne, NE4 5TG, UK}

\textit{$^2$National Quantum Computing Centre, Didcot, OX11 0QX, UK}

\textit{$^3$Yusuf Hamied Department of Chemistry, University of Cambridge, Lensfield Road, Cambridge CB2 1EW, UK}

$^*$sorayapanambalom@gmail.com

$^\dagger$jonte.hance@newcastle.ac.uk}}
\maketitle

\begin{abstract}
As quantum hardware scales to larger devices, the classical software layers that interface with it must evolve in step. Postprocessing routines developed and tested primarily in simulator settings can encode assumptions that no longer hold on utility-scale devices, leading to data loss that can be difficult to detect from high-level model outputs alone. We present a case study of \texttt{SamplerQNN}, the sampling-based quantum neural network class in the Qiskit Machine Learning library. Here, the postprocessing method applies a filter that assumes measurement bit-strings are in virtual qubit space. On our quantum hardware runs, where bit-strings span over 100 physical qubits, this filter led to the loss of 85 to 99.6\% of valid measurement shots, depending on the transpiler's qubit placement. The resulting probability vector is unnormalised, allowing distorted prediction and loss values to propagate through the model without an API-level warning. We demonstrate the impact across five experiments on two IBM backends: for inference, accuracy drops from 0.94 to 0.39 on the same raw measurements; for training, the loss signal is compressed by 22 to 27$\times$, substantially reducing the sensitivity of the optimiser to the objective landscape. The behaviour arises in all released versions of the library (0.8.4 to 0.9.0). We implemented a layout-based marginalisation fix, merged into the GitHub codebase as Pull Request \#1041, that makes \texttt{SamplerQNN} postprocessing forward-compatible with current and upcoming hardware.
\end{abstract}


\section{Introduction}
\label{sec:introduction}

Reliable software behaviour is key for reproducible scientific workflows, particularly when deploying algorithms to specialised hardware. When software assumptions fail silently, the resulting effects can be difficult to distinguish from physical noise, and can propagate into the final results and their interpretation~\cite{ferman2025errorhereexperimentalevidence}. Reproducibility studies have shown that scientific code can be difficult to execute and validate across computational environments~\cite{Brodeur2024Mass}; and a study of Harvard replication datasets found that approximately half of the code provided successfully executed~\cite{Trisovic2022}, even after automatic code-cleaning corrected simpler errors (e.g., use of absolute rather than relative paths). This has obvious impacts for reproducibility of data based on this code.

In a quantum setting, it is crucial that the classical software layers that interface with the machines and perform supporting computations are correct. This is especially important in hybrid quantum-classical settings, where the classical components are integral pieces of the computational model~\cite{Callison2022hybrid}. Most current quantum machine learning algorithms fall into this category as they are often variational, with training loops based on sampling distributions~\cite{Gokhale2026} to compute the objective value from raw counts or observables. Even outside of a hybrid setting, some classical processing is needed (e.g., collection and processing of sample data, and deciding QPU inputs). In this manuscript, we present a case study of data loss in a sampling-based postprocessing routine within the Qiskit Machine Learning (ML) library~\cite{qiskit-ml-overview}, arising when code paths originally consistent with simulator-style outputs were used with utility-scale hardware outputs. This case study highlights how classical software must evolve alongside the platforms it supports, and illustrates the impact that such issues can have on reported data.

The behaviour is traced to \texttt{SamplerQNN}, a \texttt{Sampler}-derived primitive implemented in Qiskit ML that interfaces parameterised quantum circuits to classical optimisers. It executes quantum circuits via a \texttt{Sampler} backend, collects the measurement outcomes, and postprocesses them into a probability vector that can be used for classification, regression, or other downstream tasks. The postprocessing logic was designed for simulators, which return measurement bit-strings over the virtual qubits only. However, current IBM quantum hardware behaves differently: the Qiskit transpiler maps $n$ virtual qubits onto $N$ physical qubits (where $N>100$ for Eagle and Heron devices), and the backend returns bit-strings over all $N$ measured qubits. This discrepancy between the expected and actual measurement formats causes the postprocessing step to produce biased results in hardware runs.

The postprocessing filter in \texttt{SamplerQNN.\_postprocess} discards bit-strings whose decimal integer value exceeds $2^n$. On real hardware, this eliminates 85 to 99.6\% of valid shots depending on the transpiler's qubit placement, producing a probability vector that does not sum to 1. Thus, the ML model receives an incorrect signal, but without warning the user.

This behaviour manifests itself in both training and inference. When the optimiser runs on hardware, the unnormalised probabilities propagate into the loss function at every iteration, and the resulting weight updates cannot be corrected after the fact. When the model is trained classically via a simulator and only inference is performed on hardware, the predictions are incorrect, but the trained weights are intact and the correct results can be recovered by reprocessing the raw measurements.

We identified this issue while running quantum machine learning experiments on IBM quantum hardware, traced it to the source code, and submitted a fix covering layout-based marginalisation. The issue  was reported on the Qiskit ML GitHub repository (\#1040~\cite{samplerqnn_issue1040}), and our fix was submitted as Pull Request (PR) \#1041~\cite{samplerqnn_pr1041}. The PR was reviewed by the library maintainers and merged into the official codebase on 7 May 2026. The fix will be included in the upcoming version release (installable from PyPI via \texttt{pip}); in the meantime, users can access the updated \texttt{SamplerQNN} by manually installing Qiskit ML from source\footnote{\textit{main} branch at \href{https://github.com/qiskit-community/qiskit-machine-learning}{\textcolor[HTML]{840484}{\faGithub}~{github.com/qiskit-community/qiskit-machine-learning}}.}.

We verified this behaviour in recent versions of the library, including 0.8.4 and 0.9.0. However, it was not triggered when using classical simulators because the default \texttt{Sampler} returns measurements in virtual qubit space, where the filter condition is never met. It only manifests when using real IBM quantum hardware or \texttt{SamplerV2}, with all qubits measured by default.

This paper is organised as follows. Section~\ref{sec:bug} describes the issue in detail: how the postprocessing filter works, and the two distinct failure modes we identified: one driven by qubit placement, the other by ancilla noise. Section~\ref{sec:evidence} presents the experimental evidence across five experiments on two real IBM quantum devices, showing that the issue affects both inference results (accuracy dropping from 0.94 to 0.39 on identical measurements) and the training loop (loss signal compressed by 22 to 27$\times$). Section~\ref{sec:fix} describes our fix and explains why the issue was not triggered earlier. Finally, Sections~\ref{sec:implications} and \ref{sec:conclusion} discuss the implications for Qiskit ML users, concluding with an outlook on forward-compatibility of open source software for quantum machine learning applications.

\section{The Postprocessing Data Loss Mechanism}
\label{sec:bug}

To understand the issue, it helps to recall what happens when a quantum circuit runs on real hardware. In Qiskit, the qubits in the user's circuit are called virtual qubits, while the qubits on the chip are called physical qubits.

A user defines a circuit with $n$ virtual qubits (for example, $n = 4$ for a 4-feature classifier with linear entanglement). Before execution, the circuit is transpiled: the transpiler maps each virtual qubit onto a physical qubit on the chip. On a 156-qubit backend like \texttt{ibm\_kingston}, the 4 virtual qubits might be placed at positions [0, 1, 2, 3], or equally at positions [108, 109, 110, 118], depending on the chip's connectivity graph and qubit calibration data at runtime. The remaining 152 physical qubits, called \emph{ancillas}, play no role in the computation.

After execution, the backend returns one measurement bit-string per shot, covering \emph{all} 156 physical qubits, not just the 4 virtual ones. Each bit-string is 156 bits long: the 4 bits at the virtual qubit positions carry the actual computation result, and the 152 ancilla bits are irrelevant. We note that \texttt{SamplerQNN} measures all qubits in the transpiled circuit by default. To evaluate the cost function, the postprocessing step must extract the virtual qubit bits and discard the rest.

\subsection{The Filter}

\texttt{SamplerQNN.\_postprocess} extracts the virtual qubit results using the following filter:

\begin{verbatim}
# keys -> ints, filter to valid range
for k, v in counts_i.items():
    ki = _key_to_int(k)
    if ki < 2**self.num_virtual_qubits:
        probs_i[ki] = v / total_shots
\end{verbatim}

It converts each $N$-bit measurement string to an integer and keeps it only if that integer is less than $2^n$, where $n$ is the number of virtual qubits. For $n = 4$, the threshold is $2^4 = 16$: any bit-string whose base-10 integer value is 16 or above is discarded.

This works when the virtual qubits happen to be at positions [0, 1, 2, 3], the lowest bits. In that case, a measurement where only virtual qubits are non-zero produces an integer below 16, and the filter keeps it. However, the transpiler does not guarantee low positions by default.

\subsection{Case 1: Data Loss from High Qubit Placement}

Consider a 4-qubit circuit where the transpiler places the four virtual qubits at physical positions [108, 109, 110, 118] on a 156-qubit backend. Suppose the circuit produces the virtual outcome $\ket{0110}$, meaning virtual qubit 0 measures 0, virtual qubit 1 measures 1, virtual qubit 2 measures 1, and virtual qubit 3 measures 0. On the physical chip, this sets bit 109 to 1 (virtual qubit 1) and bit 110 to 1 (virtual qubit 2), while the remaining 154 bits are ideally 0. The resulting 156-bit integer is $2^{109} + 2^{110} \approx 1.9 \times 10^{33}$, far larger than the filter threshold of $2^4 = 16$. Therefore, the filter discards this perfectly valid measurement.

In this example, the same happens for any virtual outcome except $\ket{0000}$: as soon as any virtual qubit measures $\ket{1}$, the corresponding physical bit (at position 108, 109, 110, or 118) produces an overall decimal integer far above $2^4$, and the shot is discarded. For instance

\begin{align*}
  \ket{0000}\ket{0}^{\otimes 152} &\rightarrow \text{integer } 0 < 2^4 \rightarrow \textbf{Accepted} \\
  \ket{0001}\ket{0}^{\otimes 152} &\rightarrow \text{integer } 2^{108} \gg 2^4 \rightarrow \textbf{Rejected}
\end{align*}

The only virtual outcome that \emph{can} pass the filter is $\ket{0000}$, where all four virtual qubits measure zero, and even then, only if all 152 ancilla qubits also measure zero. On real hardware, this almost never happens. In one of our experiments with this qubit placement, only 16 out of 4096 shots (0.4\%) passed the filter (see Section~\ref{sec:evidence}).

\subsection{Case 2: Data Loss from Ancilla-Bit Contributions}

The filter can also discard valid measurements even when virtual qubits are in low positions. Consider a circuit where the virtual qubits are at positions [0, 1, 2, 3]. Suppose a shot produces the virtual outcome $\ket{1010}$, a perfectly valid measurement. The integer contribution from the virtual qubits alone is $2^1 + 2^3 = 10$, which is below the threshold of 16. However, the 152 ancilla qubits are also measured, and hardware noise (thermal excitations, readout errors) causes some of them to flip. If a single ancilla at position 20 flips to $\ket{1}$, the full 156-bit integer becomes at least $2^{20} \approx 10^6$, far above the threshold. Because of this flip, the shot is discarded even though the active (i.e. non-ancilla) qubit measurement was acceptable.

Even at the lowest possible qubit positions, only 15.2\% of shots survive for the 4-qubit model. The remaining 85 to 96\% of shots carried valid virtual qubit information but were discarded because of noise on qubits that play no role in the computation.

\subsection{Consequence: Unnormalised Probabilities}

Since \texttt{total\_shots} is computed before filtering ($= 4096$) and only surviving shots contribute to the numerator, the output probabilities $p_i$ are not normalised: $\sum p_i < 1$. With a 15.2\% survival rate they, sum to $\sim$0.15; with 0.4\% survival they sum to $\sim$0.004. Therefore, the cost function is evaluated with a probability vector that is not a valid distribution.

We verified that the same filter code is present in all released versions of the library (0.8.4, 0.9.0, and the \textit{main} development branch at the time of our report). Two prior issues on the repository (\#674 and \#819) had reported problems with the output \emph{shape} of transpiled circuits, but neither identified this effect on the postprocessing step.

\section{Experimental Evidence}
\label{sec:evidence}

We discovered the issue while running five quantum machine learning experiments for a binary breast cancer classification task~\cite{panambalom2026qml} on IBM Heron devices. Each experiment used a variational classifier (either a \texttt{Sampler}-based Classifier or a \texttt{VQC}) trained on a noisy classical simulator and transferred to hardware for inference. The five experiments ran on two backends, \texttt{ibm\_kingston} (156 qubits) and \texttt{ibm\_torino} (133 qubits), depending on device availability.

The models tested are the following: a \texttt{Sampler}-based Classifier (CS), the same classifier with small-angle initialisation~\cite{grant2019initialization} and full (all-to-all) entanglement (CS SI+FE), a Variational Quantum Classifier~\cite{havlicek2019supervised} with \texttt{ZZFeatureMap} (VQC ZZ), and a Variational Quantum Classifier with \texttt{ZFeatureMap} (VQC Z). Small-angle initialisation refers to initialising all trainable parameters near zero, so the circuit starts close to the identity. Full entanglement means all-to-all qubit connectivity, which increases the number of CNOT gates. All models were trained on a noisy classical simulator using Qiskit Aer~\cite{qiskit2024, qiskit_aer} with a noise model extracted from \texttt{FakeJakartaV2}, which reproduces the noise characteristics of real IBM hardware (gate errors, readout errors, and decoherence). Training used shot-based sampling with 1024 shots per circuit evaluation. The trained weights were then transferred to real IBM hardware for inference.

Table~\ref{tab:qubit_positions} shows the physical qubit positions assigned by the transpiler for each experiment.

\begin{table}[h]
\centering
\caption{Physical qubit positions assigned by the transpiler and percentage of measurement shots surviving the \texttt{SamplerQNN} filter for each experiment.}
\label{tab:qubit_positions}
\small
\begin{tabular}{lcllc}
\hline
\textbf{Model} & \textbf{Q} & \textbf{Backend} & \textbf{Positions} & \textbf{Surv.} \\
\hline
CS              & 4 & \texttt{ibm\_kingston} & [0,1,2,3]            & 15.2\% \\
CS              & 7 & \texttt{ibm\_torino}   & [0,1,2,3,4,5,6]      & 4.1\%  \\
CS (SI+FE)      & 4 & \texttt{ibm\_kingston} & [108,109,110,118]     & 0.4\%  \\
VQC (ZZ)        & 4 & \texttt{ibm\_torino}   & [61,62,54,60]         & 0.2\%  \\
VQC (Z)         & 7 & \texttt{ibm\_torino}   & [0,1,2,3,4,5,6]      & 5.2\%  \\
\hline
\end{tabular}
\end{table}

Two experiments returned near-random accuracy (0.39), despite having trained well on the simulator. The other three transferred reasonably. The variable that cleanly separated correct results from incorrect ones was the physical qubit positions: experiments where the transpiler placed virtual qubits at positions 0 to 6 gave correct results, while those mapped to positions in the 50s or above 100 gave near-random accuracy. Since the quantum circuit performs the same computation regardless of which physical qubits it uses, this pointed to a position-dependent error in the postprocessing step.

\subsection{Impact on Inference}
\label{sec:inference}

The experiments in this section use models trained classically on a simulator. The trained weights are then used to run inference on quantum hardware, with no further optimisation on the device.
  
To confirm the impact described in Section~\ref{sec:bug}, we retrieved the raw measurement counts from all five hardware jobs and computed predictions using three different methods from the \emph{same raw data}:

\begin{itemize}
    \item \textbf{Method~A (all qubits):} compute parity over all physical qubits. This is a na\"ive baseline: the ancilla qubits are in a noisy state, so including them introduces substantial noise.
    \item \textbf{Method~B (marginalisation):} use the circuit layout to identify the physical positions of the virtual qubits, extract only those bits from each measurement, and compute parity over them. The ancilla qubits are \emph{marginalised out}. All shots contribute to the prediction.
    \item \textbf{Method~C (SamplerQNN filter):} convert each bit-string to an integer and keep only entries below $2^{n}$. This reproduces exactly what \texttt{SamplerQNN.\_postprocess} does internally.
\end{itemize}

Tables~\ref{tab:methods_acc} and~\ref{tab:methods_f1} compare the three methods across all five experiments.

\begin{table}[h]
\centering
\caption{Accuracy comparison of three postprocessing methods on the same raw hardware measurements. CS: Classifier Sampler. SI+FE: Small-Angle Initialisation with Full Entanglement. Methods A, B, and C are defined in Section~\ref{sec:inference}.}
\label{tab:methods_acc}
\small
\begin{tabular}{lcccc}
\hline
\textbf{Model} & \textbf{Qubits} & \textbf{Method A} & \textbf{Method B} & \textbf{Method C} \\
\hline
CS              & 4 & 0.57 & 0.94 & 0.94 \\
CS              & 7 & 0.63 & 0.92 & 0.89 \\
CS (SI+FE)      & 4 & 0.56 & 0.94 & 0.39 \\
VQC (ZZ)        & 4 & 0.44 & 0.54 & 0.39 \\
VQC (Z)         & 7 & 0.45 & 0.59 & 0.61 \\
\hline
\end{tabular}
\end{table}

\begin{table}[h]
\centering
\caption{F1 score (macro) comparison of three postprocessing methods on the same raw hardware measurements. CS: Classifier Sampler. SI+FE: Small-Angle Initialisation with Full Entanglement. Methods A, B, and C are defined in Section~\ref{sec:inference}.}
\label{tab:methods_f1}
\small
\begin{tabular}{lcccc}
\hline
\textbf{Model} & \textbf{Qubits} & \textbf{Method A} & \textbf{Method B} & \textbf{Method C} \\
\hline
CS              & 4 & 0.56 & 0.93 & 0.93 \\
CS              & 7 & 0.60 & 0.91 & 0.88 \\
CS (SI+FE)      & 4 & 0.56 & 0.93 & 0.37 \\
VQC (ZZ)        & 4 & 0.43 & 0.50 & 0.31 \\
VQC (Z)         & 7 & 0.45 & 0.53 & 0.57 \\
\hline
\end{tabular}
\end{table}

Method~C reproduced the original \texttt{SamplerQNN} results \emph{exactly} across all five experiments: not just accuracy, but all 35 individual classification metrics (precision, recall, and F1 per class, for both classes, across all five experiments). This confirmed that, for these experiments, the discrepancies were caused by the postprocessing filter, not by the hardware execution or our circuit design.

To illustrate the magnitude of the effect, Table~\ref{tab:sife_classif} shows the full class-wise metrics for the most affected experiment, a 4-qubit classifier whose virtual qubits were mapped to positions [108, 109, 110, 118]. Both columns are computed from the \emph{same} 4096 raw measurement shots.

\begin{table}[h]
\centering
\caption{Per-class classification metrics computed from the same raw hardware measurements. Method~C retains only 16 of 4096 shots (0.4\%); Method~B uses all 4096.}
\label{tab:sife_classif}
\small
\begin{tabular}{l ccc ccc}
\hline
 & \multicolumn{3}{c}{\textbf{Met.~C (filter)}} & \multicolumn{3}{c}{\textbf{Met.~B (marginal.)}} \\
\textbf{Class} & Prec. & Rec. & F1 & Prec. & Rec. & F1 \\
\hline
Malignant  & 0.34 & 0.74 & 0.47 & 0.95 & 0.88 & 0.91 \\
Benign     & 0.54 & 0.18 & 0.27 & 0.93 & 0.97 & 0.95 \\
\hline
Accuracy       & \multicolumn{3}{c}{0.39} & \multicolumn{3}{c}{0.94} \\
\hline
\end{tabular}
\end{table}

The classically trained model had learned well, achieving an accuracy of 0.94 when all shots were used (Method B). For this experiment, the dominant effect is postprocessing data loss: Method C discards 99.6\% of valid shots, reducing accuracy to 0.39.

The correction changes the results for two experiments significantly, and the scientific conclusions along with them. Without the correction, the Classifier Sampler with small-angle initialisation and full (all-to-all) entanglement appeared to perform poorly on hardware (accuracy$=$0.39, F1$=$0.37), and the VQC with \texttt{ZZFeatureMap} showed the same behaviour (accuracy$=$0.39, F1$=$0.31). The common factor between these two models is dense qubit connectivity: both involve many $CNOT$ gates due to full entanglement or pairwise feature interactions. Without inspecting the raw counts, this pattern could reasonably be interpreted as hardware-noise sensitivity in deeper or more entangling circuits, an explanation commonly invoked in quantum machine learning studies using near-term quantum devices.~\cite{kumar2024quantum}

After correction, the SI+FE model achieves 0.94 accuracy on \texttt{ibm\_kingston} (Table~\ref{tab:methods_acc}), identical to the standard Classifier Sampler. Despite these architectural differences, both models achieve the same accuracy after correction, showing that neither caused the poor performance. The drop to 0.39 was predominantly due to the postprocessing issue: these were simply the models where the transpiler placed qubits at high physical positions. 
The VQC with \texttt{ZZFeatureMap} shows lower accuracy on hardware after correction (0.54 with Method B) compared to the simulator (0.70), but the F1 score improves from 0.31 to 0.50, enough to change the interpretation from ``complete failure'' to ``moderate hardware degradation.''

What emerges after correction is a clearer picture: the real performance gap is between the VQC architecture and the Classifier Sampler, not between entanglement strategies. This conclusion was hidden by the postprocessing artefact.

For quantum ML applications, such as cancer classification and other clinical contexts, the difference between 0.39 and 0.94 accuracy would lead to entirely different assessments of whether a model is viable for deployment. The uncorrected results would have led to a substantially different assessment of the SI+FE model, whereas the corrected postprocessing identifies it as the best-performing configuration within this study.

\subsection{Impact on Training}

So far, the experiments have explored the effects of the postprocessing filter at inference time, i.e. the models were trained on a classical simulator and evaluated on hardware, so the issue only affected the final prediction step (sampling). The model weights, learned in this way, were unaffected, and we could recover the correct results simply by saving and reprocessing the raw measurements with marginalisation.

Here, we show that the issue has a greater impact if the training is performed \emph{on hardware}. To test this, we progressed the pre-trained models in Section~\ref{sec:inference} for 5 additional COBYLA iterations directly on IBM quantum hardware (\texttt{ibm\_torino} for 4 qubits, \texttt{ibm\_fez} for 7 qubits), using \texttt{SamplerQNN} to evaluate the loss at every iteration. This means the postprocessing data loss affects both the loss values used during the five fine-tuning evaluations and the subsequent inference step.

The loss function computes a probability-weighted expected loss as follows:

\begin{equation}
\begin{split}
\mathcal{L}&(\theta) =\\ &\frac{1}{N} \sum_{i=1}^{N} \left[ P(0 \mid x_i, \theta) \cdot (0 - y_i)^2 + P(1 \mid x_i, \theta) \cdot (1 - y_i)^2 \right]
\label{eq:loss}
\end{split}
\end{equation}

\noindent where $P(k \mid x_i, \theta)$ are the class probabilities returned by \texttt{SamplerQNN}. With the issue present and a survival rate of $\sim$4\%, the probabilities are compressed by a factor of $\sim$25, i.e. $\sum p_i\sim0.04$ instead of 1. The loss is compressed by the same factor. At iteration~1 of the on-device fine-tuning, the 4-qubit model has a correct loss of 0.33 but the optimiser received 0.012, a 27$\times$ reduction in the loss signal (Table~\ref{tab:loss_per_iter}). Similarly, for the 7-qubit model we expect a loss of 0.41, but observed a raw loss 0.019, giving a 22$\times$ reduction.

Our results show that the optimiser receives a near-constant loss signal ($\sim$0.01 to 0.02) regardless of the model's actual performance. Because of the very small gradients, it is effectively unable to suggest reliable optimisation pathways. This behaviour looks like a barren plateau~\cite{mcclean2018barren}, where vanishing gradients prevent the optimiser from learning. However, the cause here is not the circuit structure or expressibility, but the postprocessing step compressing the probability vector. This distinction matters because the symptom (near-constant loss, small gradients) is identical, and could easily be misdiagnosed as a barren plateau in practice.

Crucially, sampling loss during training is irreversible unless all sampled bit-strings are stored (which is typically impractical due to the storage overhead across many optimisation iterations). The first iteration uses the clean simulator-trained weights, but from iteration~2 onwards, COBYLA updates the weights based on the uncorrected loss values. Each subsequent evaluation may therefore use weights modified according to compressed and noisy loss estimates, allowing the effect to accumulate across the hardware fine-tuning sequence. The model parameters after 5 hardware iterations are the product of an optimiser that could not see the landscape it was navigating. 

\subsubsection{Loss Recomputation}

To measure the gap between what the optimiser received and what a correct implementation would have provided, we retrieved the raw measurement data from all 10 optimisation jobs (5 iterations $\times$ 2 models) and recomputed the loss using the three methods. Table~\ref{tab:loss_per_iter} shows the results.

\begin{table}[t]
\centering
\caption{Loss recomputed from raw measurements at each hardware iteration. Method~C is what the optimiser received; Method~B is the correct loss. Iteration~1 uses clean simulator-trained weights; iterations 2 to 5 use weights affected by prior uncorrected updates.}
\label{tab:loss_per_iter}
\small
\begin{tabular}{cl ccc c}
\hline
\textbf{Model} & \textbf{Iter.} & \textbf{A} & \textbf{B} & \textbf{C} & \textbf{Surv.} \\
\hline
\multirow{5}{*}{4Q}
& 1 (clean)     & 0.50 & 0.33 & 0.012 & 3.7\% \\
& 2 (affected) & 0.50 & 0.42 & 0.016 & 3.7\% \\
& 3 (affected) & 0.50 & 0.33 & 0.012 & 3.7\% \\
& 4 (affected) & 0.50 & 0.41 & 0.016 & 4.0\% \\
& 5 (affected) & 0.50 & 0.34 & 0.014 & 4.0\% \\
\hline
\multirow{5}{*}{7Q}
& 1 (clean)     & 0.50 & 0.41 & 0.019 & 4.6\% \\
& 2 (affected) & 0.50 & 0.42 & 0.019 & 4.4\% \\
& 3 (affected) & 0.50 & 0.43 & 0.018 & 4.2\% \\
& 4 (affected) & 0.50 & 0.44 & 0.015 & 3.4\% \\
& 5 (affected) & 0.50 & 0.44 & 0.015 & 3.4\% \\
\hline
\end{tabular}
\end{table}

\begin{figure*}[t]
\centering
\includegraphics[width=\textwidth]{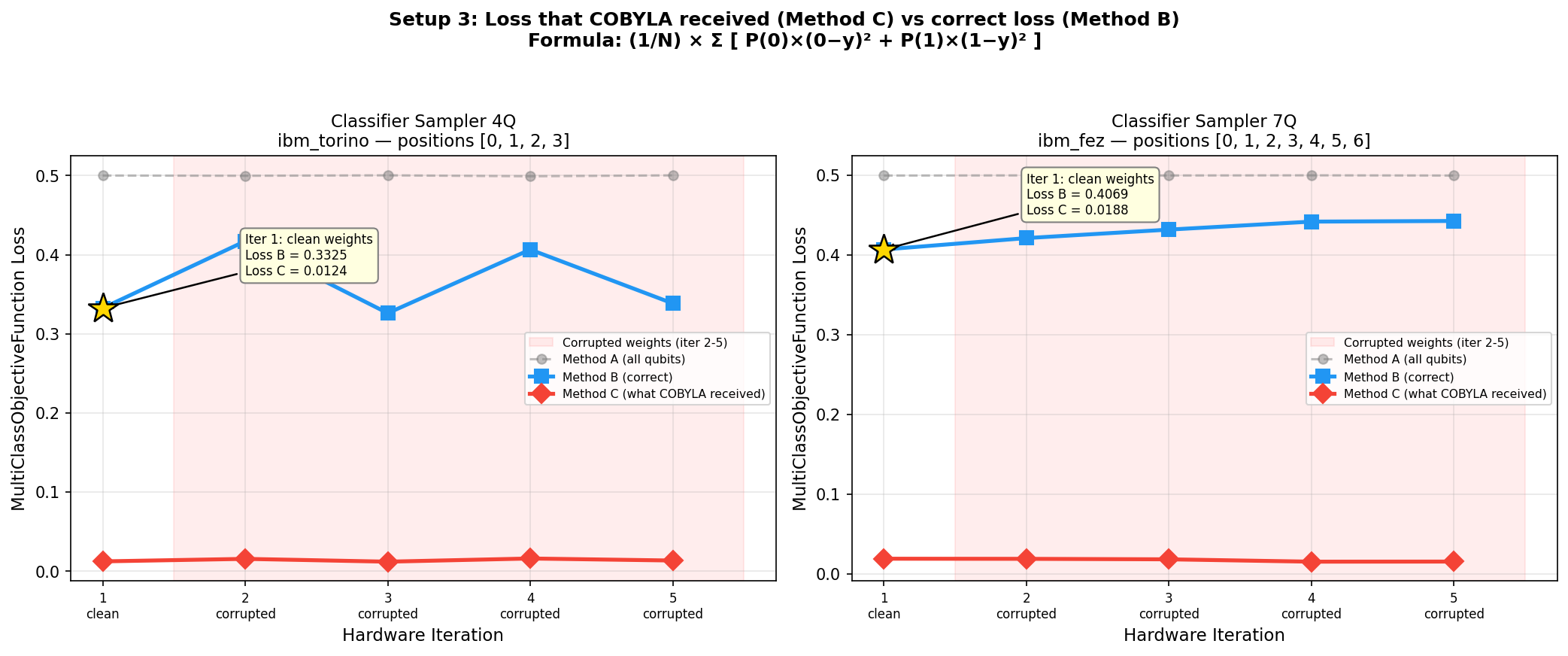}
\caption{Loss for 5 on-device fine-tuning iterations, recomputed from raw measurements. Method~B (correct marginalisation) reveals the actual loss landscape; Method~C (original filter) shows what COBYLA received. The star marks iteration~1 (clean simulator-trained weights); the shaded region marks iterations where the weights had already been affected by prior uncorrected updates.}
\label{fig:loss_comparison}
\end{figure*}

Figure~\ref{fig:loss_comparison} shows the comparison visually.

The loss in Method~C remains approximately constant between 0.01 and 0.02 throughout the fine-tuning, while Method~B shows the fiducial loss values. The loss gap between Methods~B and C persists and is large: the optimiser was navigating a landscape it could not see.

\subsubsection{COBYLA and the Simplex Constraint}

Not only was the optimiser blind, it never even began optimising. COBYLA~\cite{powell1994cobyla} is a trust-region-based (gradient-free) optimiser that works by maintaining a simplex: a set of $n+1$ points in the $n$-dimensional parameter space, where $n$ is the number of trainable weights. It uses these points to build a local linear model of the loss function. But before it can take a single optimisation step, it must first \emph{construct} this simplex by evaluating the loss at $n+1$ different parameter settings: the starting point, plus $n$ perturbations where one weight at a time is shifted by $+1.0$~rad.

The 4-qubit model has 12 trainable weights, so the simplex requires 13 evaluations. The 7-qubit model has 21 weights, requiring 22 evaluations. With only 5 hardware iterations available, COBYLA never completes its initial simplex for either model: it builds 5 out of 13 vertices for the 4-qubit model and 5 out of 22 for the 7-qubit model. Zero actual optimisation steps are taken. Every iteration we observed was part of the simplex construction phase~\cite{powell1994cobyla}.

This means the experiment was structurally under-dimensioned regardless of the issue: even if \texttt{SamplerQNN} had worked correctly and COBYLA had received accurate loss values, 5 iterations would still have been insufficient for a single optimisation step. This imposed a practical runtime constraint: each iteration takes $\sim$8~minutes of hardware time, and completing the simplex alone would require $13 \times 8 = 104$~minutes (4 qubits) or $22 \times 8 = 176$~minutes (7 qubits) out of our total quantum computing hardware allocation of 180~minutes.

\subsubsection{Weight Drift Analysis}

To verify that the iterations correspond to simplex construction rather than optimisation, we extracted the trainable weights at each iteration and compared them to the clean simulator-trained values. Table~\ref{tab:weight_drift} shows these results.

\begin{table*}[t]
\centering
\caption{Weight drift relative to clean simulator-trained values at each iteration. All non-zero changes are exactly $\pm$1.0~rad (= \texttt{rhobeg}), confirming simplex probing. Loss~C is the value the optimiser received and used to decide whether to keep or revert each perturbation.}
\label{tab:weight_drift}
\small
\begin{tabular}{cl cccc c c}
\hline
\textbf{Model} & \textbf{Iter.} & $\Delta w_0$ & $\Delta w_1$ & $\Delta w_2$ & $\Delta w_3$ & \textbf{\# changed} & \textbf{Loss~C} \\
\hline
\multirow{6}{*}{4Q (12 weights)}
& 1 (clean)     & 0 & 0 & 0 & 0 & 0 & 0.012 \\
& 2 & +1.0 & 0 & 0 & 0 & 1 & 0.016 \\
& 3 & 0 & +1.0 & 0 & 0 & 1 & 0.012 \\
& 4 & 0 & +1.0 & +1.0 & 0 & 2 & 0.016 \\
& 5 & 0 & +1.0 & 0 & +1.0 & 2 & 0.014 \\
& inference & 0 & +1.0 & 0 & 0 & 1 & $-$ \\
\hline
\multirow{6}{*}{7Q (21 weights)}
& 1 (clean)     & 0 & 0 & 0 & 0 & 0 & 0.019 \\
& 2 & +1.0 & 0 & 0 & 0 & 1 & 0.019 \\
& 3 & +1.0 & +1.0 & 0 & 0 & 2 & 0.018 \\
& 4 & +1.0 & +1.0 & +1.0 & 0 & 3 & 0.015 \\
& 5 & +1.0 & +1.0 & +1.0 & +1.0 & 4 & 0.015 \\
& inference & +1.0 & +1.0 & +1.0 & +1.0 & 4 & $-$ \\
\hline
\end{tabular}
\end{table*}

Every perturbation is exactly $+1.0$~radian along a single parameter axis, consistent with simplex construction, not optimisation. Whether COBYLA keeps or reverts a perturbation depends on whether the perturbed loss appears better or worse than the reference. Without the issue, these decisions would have been informed by the real loss values (Method~B), which show clear differences between iterations. With the issue, COBYLA received compressed values of $\sim$0.01 to 0.02 where the differences ($\sim$0.001 to 0.003) are dominated by shot noise. In this regime, the keep-or-revert decisions are effectively random.

For the 4-qubit model, only 1 of 4 perturbations survives to inference, and the model stays near the simulator optimum. For the 7-qubit model, all 4 perturbations are kept, and the correct loss rises monotonically from 0.41 to 0.44, and each accumulated perturbation pushes the model further from the optimum.

\section{Layout-based marginalisation}
\label{sec:fix}

A hardware-compatible postprocessing implementation requires layout-aware marginalisation, outlined as follows. For each measurement bit-string, extract only the bits at the physical positions corresponding to virtual qubits, reconstruct an integer in the virtual qubit space, and accumulate probabilities. This is equivalent to summing over all possible ancilla states, the standard mathematical operation in quantum mechanics for projecting a joint distribution onto a subspace. It uses every shot, is unaffected by ancilla noise, and produces a probability distribution that sums to 1.0 by construction.

We reported the issue on the Qiskit ML repository (issue \#1040~\cite{samplerqnn_issue1040}), together with the full diagnostic: the qubit position pattern across five experiments, the three postprocessing methods with accuracy and F1 comparisons, the per-class classification reports confirming exact reproduction of the original results, and a proposed fix based on marginalisation. The maintainers flagged the report as high priority and invited a PR implementing the proposed fix.

Our PR~\cite{samplerqnn_pr1041} replaces the integer value filter in \texttt{\_postprocess} with layout-based marginalisation as follows:

\begin{verbatim}
pos = layout.final_index_layout()
for key, count in counts.items():
    key_int = to_integer(key)
    # extract bit at each virtual position
    bits = [(key_int >> q) & 1 for q in pos]
    # reconstruct virtual-space integer
    virtual = sum(
        b << i for i, b in enumerate(bits)
    )
    probs[virtual] += count / total_shots
\end{verbatim} 

When a circuit layout is present (i.e., the circuit has been transpiled), the fix reads the physical positions of the virtual qubits from the layout, extracts only those bits from each measurement bit-string via bit-shifting, and accumulates probabilities in the virtual qubit space. Shots that differ only in their ancilla bits are mapped to the same virtual key, effectively marginalising the ancillas out. When no layout is present (local simulation with the default sampler), the original behaviour is preserved, as no ancilla qubits exist in that case and the filter works correctly. The PR was reviewed and merged into the official codebase on 7 May 2026, and will be distributed on PyPI in the next version release.

The issue had not been reported before because the default sampler used by \texttt{SamplerQNN} (the internal \texttt{QMLSampler}, based on state vector simulation) returns bit-strings in \emph{virtual} qubit space: $n$-bit strings rather than $N_{\text{physical}}$-bit strings. In that case, the integer value of any measurement is at most $2^n - 1$, so the filter threshold is never exceeded. The problem only manifests when using real hardware or \texttt{SamplerV2}, which return measurements over all physical qubits. Since most tutorials and examples in the Qiskit ecosystem use simulators, and real hardware access requires specific allocation programmes, the issue went undetected despite being present in all released versions of the library (including 0.8.4 and 0.9.0 at the time of our report).

Moreover, the filter may not have caused issues on earlier IBM quantum devices, which had approximately 30 qubits or less~\cite{havlicek2019supervised, mensa2023quantum}, making it more likely for virtual qubits to be mapped to low-numbered physical positions that would pass the threshold. As hardware evolved toward larger physical qubit registers, like the Eagle and Heron IBM architectures with over 100 qubits, the same assumption became more likely to induce substantial data loss.

\section{Broader Implications}
\label{sec:implications}

To assess the potential scope of the behaviour, we searched for published studies whose workflows match the conditions under which the postprocessing data loss can occur: \texttt{SamplerQNN} usage, the Qiskit ML versions we tested, and execution on real quantum hardware or \texttt{SamplerV2}-like outputs. We discuss two examples of such works below.

Martin-Perez et al.~\cite{martinperez2026qtl} proposed a hybrid classical-quantum transfer learning pipeline in which a pre-trained convolutional neural network (CNN) backbone (ResNet18, EfficientNet-B0, or MobileNetV2) extracts image features, and a \texttt{SamplerQNN} with 4 qubits replaces the final classifier. They used Qiskit ML 0.8.0 and ran hardware experiments on \texttt{ibm\_torino} (133 qubits) using \texttt{SamplerV2}. Their results showed that one backbone (ResNet18) slightly improves on hardware compared to noisy simulation (+1.9\%), while another (EfficientNet-B0) drops by 19 percentage points (from 91.03\% to 71.79\%). The authors attributed this gap to ``transpilation overhead and stochastic gradients''~\cite{martinperez2026qtl}. Because the same quantum circuit was used for all three backbones, the selective drop is compatible with several explanations, including hardware noise, input-dependent sampling effects, and the postprocessing data-loss mechanism described in our work. We note that we cannot confirm whether the behaviour we document caused the gap in Ref.~\cite{martinperez2026qtl} without inspecting their raw measurement data, but the setup (4 virtual qubits on a 133-qubit backend via \texttt{SamplerQNN} and \texttt{SamplerV2}) matches the conditions under which the behaviour arises.

Chaudhary et al.~\cite{chaudhary2026qagnn} proposed a quantum-enhanced graph neural network for intrusion detection. Their methodology describes expectation values of Pauli observables (consistent with \texttt{EstimatorQNN}), and their simulation experiments use \texttt{StatevectorEstimator} and \texttt{BackendEstimatorV2}. However, for their hardware validation on \texttt{ibm\_fez} (156 qubits), they switch to \texttt{SamplerQNN} with 128 shots and Qiskit ML 0.9.0. With 4 virtual qubits on a 156-qubit backend, this workflow satisfies the conditions under which the original range filter can discard a large fraction of valid measured shots, depending on layout and measurement-register details. In this case, the impact of data loss is not directly measurable: the hardware experiment uses a very small dataset (8 training nodes, 5 testing nodes), and both the hardware and the noisy simulator produce flat loss curves over 30 epochs. As with Ref.~\cite{martinperez2026qtl}, we cannot confirm the issue affected their results without access to the raw measurement data, however, the setup matches the conditions for potential data loss in postprocessing.

More broadly, performance degradation when moving from simulation to real hardware is common in quantum machine learning and is typically attributed to hardware noise. Our findings show that software-level errors in the postprocessing pipeline can produce a similar effect, and the two are difficult to distinguish without determining whether raw bit-string counts are normalised. For previously published results obtained with affected versions (0.8.4 to 0.9.0), correct inference results can still be recovered if the raw measurement counts were saved, by applying marginalisation as described in Section~\ref{sec:fix}. Training results cannot be corrected after the fact, since the uncorrected loss signal has already been applied to the model weights at each optimisation step.
It is worth noting that only \texttt{SamplerQNN}-based workflows may trigger data loss under the conditions above. Other Qiskit ML tools, like quantum kernels or Bayesian inference are not expected to suffer data loss of the same nature, as their internal workflows never call \texttt{SamplerQNN}.

\section{Conclusion}
\label{sec:conclusion}

We analysed and corrected a postprocessing data-loss mechanism in \texttt{SamplerQNN} that, in the hardware experiments we performed, discarded 85 to 99.6\% of valid measurement shots. We verified the behaviour in the affected released versions examined here, including Qiskit ML 0.8.4 and 0.9.0, and found the root cause to a filter that assumed bit-strings are in virtual qubit space. This assumption holds for classical simulators but not for real hardware backends where bit-strings span large (over 100) physical qubit registers in the device.

Our experiments showed that the issue affects both inference and training. For inference, it can reduce accuracy from 0.94 to 0.39 on identical raw measurements. For training, it degrades the cost function evaluation by 22 to 27$\times$, preventing the classical optimiser from making meaningful updates.

Our fix, based on layout marginalisation, was merged into the official repository (PR \#1041) and will be available on PyPI from version 0.9.1. Inference results from previous Qiskit ML versions can be corrected if the raw measurement counts were saved; however, models trained on hardware cannot be corrected \textit{post hoc} and require re-running with the fix.
While researchers routinely validate their own workflows, this case study shows that assumptions at the boundary between user code, library code, transpilation, and hardware outputs also require validation as platforms evolve. Software-level explanations should be considered alongside hardware noise, model expressibility, and optimisation artifacts when interpreting unexpected experimental results. This is especially crucial as software scope and complexity grows with algorithmic and hardware advances.
Open-source quantum computing frameworks have been instrumental in making quantum hardware accessible to researchers; maintaining their reliability as hardware scales requires close feedback between users, maintainers, and hardware providers, and benefits the entire community.

\section*{Acknowledgment}
The authors thank Laura Martin for her help with data recovery. They also thank the Qiskit ML maintainers who reviewed and merged the fix.
SP acknowledges support from the UKRI National Edge AI Hub for Real Data: Edge Intelligence for Cyber-disturbances and Data Quality (EPSRC EP/Y028813/1).
NC and JRH acknowledge support from their EPSRC Mathematical Sciences Small Grant (UKRI3647).
NC acknowledges support from QCI3 - the Hub for Quantum Computing via Integrated and Interconnected Implementations (EP/Z53318X/1).
JRH acknowledges support from a Royal Society Research Grant (RG/R1/251590), and from their EPSRC Quantum Technologies Career Acceleration Fellowship (UKRI1217).
This project was funded and supported by the UK National Quantum Computer Centre (NQCC) [NQCC200921], which is a UKRI Centre and part of the UK National Quantum Technologies Programme (NQTP). Access to quantum processing units was enabled through the NQCC SparQ programme, with additional compute time provided through the NQCC's in-kind contribution to JRH's EPSRC Quantum Technologies Career Acceleration Fellowship.

\section*{Data and Code Accessibility}
All data and code supporting this work can be accessed from \href{https://www.doi.org/10.5281/zenodo.22304590}{doi:10.5281/zenodo.22304590}.

\bibliographystyle{IEEEtran}
\bibliography{references}

\end{document}